# Origin of Quantum Noise and Decoherence in Distributed Amplifiers

J.D. Franson and B.T. Kirby

*Physics Department, University of Maryland at Baltimore County, Baltimore, MD 21250*

The use of distributed amplifiers may have some potential advantages for the transmission of quantum information through optical fibers. In addition to the quantum noise introduced by the amplifiers, entanglement between atoms in the amplifying media and the optical field corresponds to which-path information that can further reduce the coherence. Here we analyze the effects of decoherence in a phase-insensitive distributed amplifier by using perturbation theory to calculate the state of the entire system including the atomic media. For an initial coherent state, tracing over the atomic states allows the reduced density matrix of the field to be expressed as a mixture of squeezed states with a reduced spread in photon number and an increased phase uncertainty. The amplifier noise and decoherence can be interpreted as being due to entanglement with the environment rather than the amplification of vacuum fluctuation noise.

**I. INTRODUCTION**

The transmission of quantum information over large distances in optical fibers is limited by the effects of loss and decoherence. It has previously been suggested that the use of distributed amplifiers may have some potential benefits in reducing the rate of decoherence [1-5]. Roughly speaking, noise that is introduced by amplification in one part of the fiber will be attenuated by loss in subsequent parts of the fiber. In addition, the total power dissipated in the amplifiers is less than it would be for a single amplifier placed at the beginning of the system, for example, which reduces the amount of which-path information left in the environment. Here we analyze the noise and decoherence produced by a distributed amplifier system including the effects of entanglement with the environment.

The quantum noise produced by an optical amplifier has been investigated since the earliest days of quantum optics [6-11]. Most of those calculations were based on the introduction of a noise operator as required by unitarity [4,6,8] or on the master equation and related techniques [2,10,11]. For a linear amplifier, this results in a quantum noise that is added to the signal, and calculations of that kind characterize the statistical properties of the added noise.

But in addition to the quantum noise in the signal, decoherence can also occur as a result of entanglement between the amplifying medium and the optical field. This can be viewed as which-path information that can partially or completely distinguish between the two components of a Schrodinger cat state, for example. Thus it is necessary to explicitly include the entanglement with the environment, and the results in general are not equivalent to an additive noise. Earlier papers on the effects of amplification on Schrodinger cat states [12-17] did not consider the important case of a distributed amplifier.

We consider a distributed phase-insensitive amplifier system in the limit in which the average intensity of the signal is never allowed to increase or decrease by a significant amount. Instead, a small amount of amplification is assumed to alternate with a small amount of loss in order to maintain a nearly constant intensity throughout the length of the optical fiber or other transmission channel. This allows the use of perturbation theory to calculate the quantum state of the entire system, including the environment (atoms in the loss and amplifying media) as well as the electromagnetic field.

We initially consider the propagation of a coherent state $|\alpha\rangle$ through such a system and later generalize the approach to superposition states as well. By tracing over the state of the atoms, the reduced density matrix of the field is shown to be equivalent to a mixture of squeezed coherent states $|\tilde{\alpha}\rangle$, each of which has a reduced spread in photon number and an increased phase uncertainty. The nature of the mixed state also increases the overall uncertainty in the photon number, so that the system is no longer in a minimum uncertainty state. These results suggest that amplifier noise and decoherence can both be interpreted as being due to entanglement with the environment rather than the amplification of vacuum fluctuations.

The remainder of the paper begins in section II with a simple model for a phase-insensitive amplifier based on a series of interactions with individual atoms. Perturbation theory is used in Section III to calculate the state of the total system and to determine the probability distributions for the photon number and the number of atoms that have made a transition to a different state. The reduced density matrix of the field is calculated in Section IV by tracing over all of the atomic states. Section V shows that the reduced density matrix can be written in an equivalent but more useful form as a mixture of squeezed coherent states $|\tilde{\alpha}\rangle$. Section VI uses the coordinate representation of the field to calculate the probability distribution of the phase of the field as measured using homodyne techniques. Section VII applies these results to a superposition of coherent states (Schrodinger cat) while a summary and conclusions are presented in Section VIII.

## II. SIMPLE MODEL OF A PHASE-INSENSITIVE DISTRIBUTED AMPLIFIER

In a distributed amplifier, the loss in an optical fiber or other transmission line is compensated by inserting amplifiers into the optical path at frequent intervals as illustrated in Fig 1a. We will consider the limiting case in which the number $N_S$ of amplifier sections is sufficiently large that the loss is compensated nearly continuously. This prevents the intensity of the signal from varying significantly from its initial value.

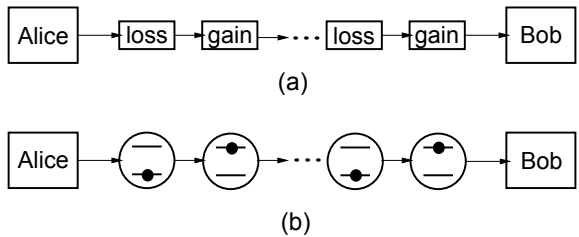

Fig. 1. (a) A phase-insensitive distributed amplifier system with alternating sections of loss and amplification. This becomes a quasi-continuous process in the limit of a large number $N_S$ of sections with small loss and amplification. (b) As $N_S$ goes to infinity, this is equivalent to a weak interaction with a single atom in its ground state that produces a small amount of loss, followed by an interaction with a single atom in its excited state (inverted population) to give the corresponding amplification or gain. Repeating this process many times gives a simple model for a continuous phase-insensitive distributed amplifier system.

In the limit as $N_S \to \infty$, the mean number of photons lost in a short section of fiber and the mean number of photons regenerated by a single amplifier are both much less than one. In that case the system is equivalent to an interaction with a single two-level atom in each successive section of the amplifier or fiber as illustrated in Fig. 1b. The atoms associated with the loss mechanism are assumed to be in their ground state initially while the atoms in the amplifiers are assumed to initially be in their excited level (a population inversion). The loss could also be modeled by a series of beam splitters, but it is more convenient here to represent both the amplifiers and the loss mechanism in a similar way.

This model of a distributed phase-insensitive amplifier is equivalent to having a continuous distribution of amplifier atoms throughout the loss medium provided that the atoms have a negligible coherence time. The assumption that the optical signal only interacts once with any given atom avoids any coherent atomic effects in which entanglement (which-path information) is subsequently reduced by Rabi oscillations, for example. As a result, the model considered here provides a conservative estimate of the amount of coherence left in the optical signal.

We will first consider the effects of a continuously distributed amplifier of this kind on an quasi-monochromatic incident coherent state $|\alpha\rangle$, which is defined by [18]

$$|\alpha\rangle = e^{-\alpha^*\alpha/2} \sum_{n=0}^{\infty} \frac{\alpha^n}{\sqrt{n!}} |n\rangle \qquad (1)$$

where $|n\rangle$ is a state containing $n$ photons. In the limit of large $|\alpha|$, Stirling's approximation can be used to rewrite Eq. (1) in the form [19]

$$|\alpha\rangle = \frac{1}{(2\pi n_0)^{1/4}} \sum_{n=0}^{\infty} e^{in\phi} e^{-(n-n_0)^2/2\sigma_n^2} |n\rangle \equiv \sum_{n=0}^{\infty} c_n |n\rangle. \qquad (2)$$

Here the phase $\phi$ is defined by $\alpha = |\alpha|e^{i\phi}$, $n_0 = \alpha^*\alpha$ is the mean number of photons, and $\sigma_n = \sqrt{2n_0}$ is the width of the photon number probability amplitude distribution. (The standard deviation of the photon number probability distribution is a factor of $\sqrt{2}$ smaller.) This corresponds to a Gaussian distribution for the photon number as illustrated in Fig. 2, where the constant $c_n$ is defined by Eq. (2).

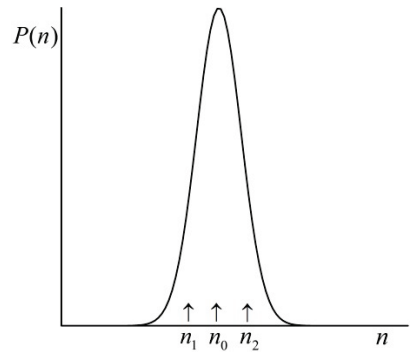

Fig. 2. Probability distribution $P(n)$ for the number $n$ of photons in a coherent state $|\alpha\rangle$. The mean photon number is $n_0 = |\alpha|^2$, while $n_1$ and $n_2$ are two other typical values whose effects on the environment are different, as illustrated in Fig. 3. (Arbitrary units.)

We will denote the total number of loss atoms in Fig. 1 by $N_T$, which is also equal to the total number of amplifier atoms. The state $|\psi_0\rangle$ of the system before any interaction is then given by

$$|\psi_0\rangle = |\alpha\rangle \prod_{i,j=1}^{N_T} |G_{Li}\rangle |E_{Aj}\rangle = \sum_{n=0}^{\infty} c_n |n\rangle \prod_{i,j=1}^{N_T} |G_{Li}\rangle |E_{Aj}\rangle \qquad (3)$$

Here $|G_{Li}\rangle$ represents the ground state of loss atom $i$ while $|E_{Aj}\rangle$ represents the excited state of amplifier atom $j$.

The interaction Hamiltonian $\hat{H}'$ of the system can be chosen as usual to have the form

$$\hat{H}' = i\varepsilon \sum_i g_i(t)\left(\hat{\sigma}_i^+ \hat{a} + \hat{\sigma}_i^- \hat{a}^\dagger\right) \\ + i\varepsilon \sum_j g_j(t)\left(\hat{\sigma}_j^+ \hat{a} + \hat{\sigma}_j^- \hat{a}^\dagger\right). \quad (4)$$

Here $\varepsilon \ll 1$ is a constant related to the atomic matrix elements, the operators $\hat{\sigma}_i^+$ and $\hat{\sigma}_i^-$ raise or lower the state of atom $i$, while $\hat{a}^\dagger$ and $\hat{a}$ create and annihilate a photon. The time-dependent factors of $g_i(t)$ and $g_j(t)$ represent the fact that the various atoms are sequentially coupled to the field. The arbitrary phases of the atomic eigenstates have been chosen in such a way that $\varepsilon$ is a real number.

### III. PERTURBATION THEORY

Because the individual interactions are weak, perturbation theory can be used to calculate the state of the system $|\psi_1\rangle$ after the coherent state has passed through the first absorbing atom. If we assume that all of the atoms are on resonance with the optical field and that the interaction occurs ($g_{i=1}(t) = 1$) over a time interval of $\Delta t$, then

$$|\psi_1\rangle = \sum_{n=0}^{\infty}\left(1 - \frac{\varepsilon^2 n \Delta t^2}{\hbar^2}\right) c_n |n\rangle \prod_{i,j}^{N_T} |G_{Li}\rangle |E_{Aj}\rangle \\ + \sum_{n=1}^{\infty}\left(\frac{\varepsilon n^{1/2} \Delta t}{\hbar}\right) c_n |n-1\rangle |E_{L1}\rangle \prod_{i>1,j}^{N_T} |G_{Li}\rangle |E_{Aj}\rangle. \quad (5)$$

It can be seen that there is a probability $P_n = \varepsilon^2 n \Delta t^2 / \hbar^2$ that the number state component $|n\rangle$ will be reduced to $|n-1\rangle$ by the absorption of a photon.

The state of the system after the subsequent interaction with the first amplifier atom can be determined in the same way. This gives a probability $P'_n = \varepsilon^2 (n+1) \Delta t^2 / \hbar^2$ that the number state component $|n\rangle$ will be increased to $|n+1\rangle$ by the emission of a photon, with $P'_n \doteq P_n$ in the limit of large $n_0$ since $\sqrt{n+1} \doteq \sqrt{n}$ in the relevant matrix elements.

Since there is a probability $P_n$ of increasing or decreasing the photon number at each step in the process, the photon number distribution will undergo an unbiased random walk as illustrated in Fig. 3a. A term in the initial coherent state that initially had exactly $n$ photons will now have a probability distribution $P_p(n';n)$ that it will contain $n'$ photons at the end of the process. From the central limit theorem [20], $P_p(n';n)$ will approach a Gaussian distribution in the limit of a large number of absorption and emission events as illustrated schematically in Fig. 3a.

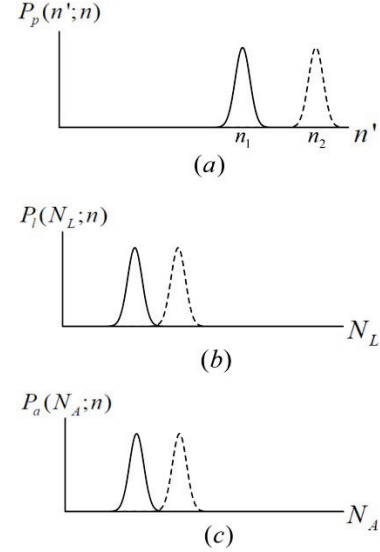

Fig. 3. Effects of the random walk process produced by the absorption and emission of photons. (a) Probability distribution $P_p(n';n)$ for the final number of photons, where the solid line is the distribution resulting from $n = n_1$ initial photons while the dashed line is the distribution resulting from $n = n_2$ initial photons. (b) The probability distribution $P_l(N_L;n)$ for the final number $N_L$ of loss atoms that were transferred to their excited state, where the solid line corresponds once again to $n = n_1$ initial photons while the dashed line corresponds to $n = n_2$ initial photons. (c) The corresponding probability distribution $P_a(N_A;n)$ for the final number $N_A$ of amplifier atoms that were transferred to their ground state. The point is that different initial photon number states can lead to nearly orthogonal states of the environment. (Arbitrary units.)

The probability distributions for the number of atoms that have made a transition to a different state at the end of the process will also be of interest. The total number of loss atoms that have made a transition to their excited state at the end of the transmission process will be denoted by $N_L$, while the total number of amplifier atoms that made a transition to their ground state will be denoted by $N_A$. The probability distribution for $N_L$ will be denoted by $P_l(N_L;n)$, which is a function of the initial number $n$ of photons in the field. Similarly, $P_a(N_A;n)$ will denote the probability distribution for $N_A$ given $n$ photons initially in the field. These distributions will also become Gaussian distributions in the limit of a large number of emission and absorption events as illustrated in Figs 3b and 3c.

$N_L$ is described by a random process in which there is a small probability $P_n$ that $N_L$ will increase by one at each step. This produces a Poisson probability distribution for $P_l(N_L;n)$ that is given by [20]

$$P_l(N_L;n) = \frac{e^{-\mu(n)}\mu^{N_L}}{N_L!}. \quad (6)$$

Here the mean value of $N_L$ is $\mu(n) = N_T P_n$ and the standard deviation is $\sigma_L = \sqrt{\mu(n)}$. From the central limit theorem [20] or Stirling's approximation [19], Eq. (6) reduces in the limit of large $N_L$ to a normal distribution given by

$$P_l(N_L;n) \rightarrow f(N_L;n) = c_L e^{-(N_L-\mu(n))^2/2\sigma_L^2}, \quad (7)$$

where $c_L = 1/(\sqrt{2\pi}\sigma_L)$ is a normalizing constant. For large values of the initial mean photon number $n_0$, $\sigma_L$ and $c_L$ are approximately independent of the value of $n$.

The probability distribution $P_a(N_A;n)$ for $N_A$ also reduces to $f(N_A;n)$ as given by Eq. (7) in the limit of large $n_0$ where $\sqrt{n+1} \doteq \sqrt{n}$. From conservation of energy, the total change $\Delta n$ in the number of photons is given by

$$\Delta n = N_A - N_L. \quad (8)$$

The atomic distributions $f(N_L;n)$ and $f(N_A;n)$ can be combined with Eq. (6) to show that the probability distribution $f_p(\Delta n)$ for $\Delta n$ is given by

$$f_p(\Delta n) = \frac{1}{\sqrt{2\pi}\sigma_{\Delta n}} e^{-\Delta n^2/2\sigma_{\Delta n}^2} \quad (9)$$

where $\sigma_{\Delta n} = \sqrt{2}\sigma_L$.

It can be seen from Eqs. (4) and (5) that all of the possible terms in the final quantum state of the system will be generated with the same phase. Their probability amplitudes are given by the square-root of the corresponding probabilities. As a result, the final state $|\psi'\rangle$ of the system after the interaction with $N_T$ loss and amplifier atoms can be written in the form

$$|\psi'\rangle = \sum_n c_n \sum_{N_A} \sum_{N_L} f^{1/2}(N_L;n) f^{1/2}(N_A;n) \\ \times |n+\Delta n\rangle|N_L\rangle|N_A\rangle. \quad (10)$$

Here $\Delta n = N_A - N_L$ from Eq. (8) while $|N_L\rangle$ and $|N_A\rangle$ denote states of the environment in which $N_L$ loss atoms are left in their excited states and $N_L$ amplifier atoms are left in their ground state.

### IV. DENSITY MATRIX

The width $\sigma_L$ of the distributions $f(N_L;n)$ and $f(N_A;n)$ only increases as $\sqrt{\mu(n)}$, which means that their relative width $\sigma_L/\mu(n)$ will decrease in the limit of a large number of interactions. As a result, the distributions associated with two different photon number components $|n_1\rangle$ and $|n_2\rangle$ in the initial coherent state will have negligible overlap if $n_1$ and $n_2$ differ by a sufficiently large amount, as illustrated in Figs. 3b and 3c. This effect will eliminate any coherence between terms with sufficiently large differences in initial photon number, which produces a mixed state with an increased phase uncertainty as we will now show.

Eq. (10) is a pure state with a density operator $\hat{\rho}$ given by

$$\hat{\rho} = \sum_{n,n'} c_n c_{n'}{}^* \sum_{N_A,N_{A'},N_L,N_{L'}} f^{1/2}(N_L;n) f^{1/2}(N_A;n) f^{1/2}(N_{L'};n') \\ \times f^{1/2}(N_{A'};n')|n+\Delta n\rangle|N_L\rangle|N_A\rangle\langle N_{A'}|\langle N_{L'}|\langle n'+\Delta n'|. \quad (11)$$

This can be rewritten as a sum over $\Delta n$ and $\Delta n'$ using the fact that $N_A = N_L + \Delta n$ and $N_{A'} = N_{L'} + \Delta n'$ to obtain

$$\hat{\rho} = \sum_{n,n'} c_n c_{n'}{}^* \sum_{N_L,N_{L'}} \sum_{\Delta n,\Delta n'} f^{1/2}(N_L;n) f^{1/2}(N_L+\Delta n;n) \\ f^{1/2}(N_{L'};n') f^{1/2}(N_{L'}+\Delta n';n')|n+\Delta n\rangle|N_L\rangle|N_L+\Delta n\rangle \\ \times \langle N_{L'}+\Delta n'|\langle N_{L'}|\langle n'+\Delta n'|. \quad (12)$$

The reduced density operator $\hat{\rho}_R$ for the field alone can be found by taking the partial trace of Eq. (12) over the state of the environment, which gives

$$\hat{\rho}_R = \sum_{n,n'} \sum_{\Delta n} \sum_{N_L} c_n c_{n'}{}^* f^{1/2}(N_L;n) f^{1/2}(N_L;n') \\ \times f^{1/2}(N_L+\Delta n;n) f^{1/2}(N_L+\Delta n;n')|n+\Delta n\rangle\langle n'+\Delta n|. \quad (13)$$

Here we have made use of the fact that two states of the environment will be orthogonal unless $N_L' = N_L$ and $N_A' = N_A$. This is equivalent to requiring that $N_L' = N_L$ and $\Delta n' = \Delta n$.

It will be convenient to define the environmental overlap $I(n,n';\Delta n)$ as

$$I(n,n';\Delta n) = \sum_{N_L} f^{1/2}(N_L,n) f^{1/2}(N_L,n') \quad (14)$$
$$\times f^{1/2}(N_L + \Delta n, n) f^{1/2}(N_L + \Delta n, n').$$

The reduced density operator can then be rewritten as

$$\hat{\rho}_R = \sum_{n,n'} c_n c_{n'}^* \sum_{\Delta n} I(n,n';\Delta n) |n+\Delta n\rangle\langle n'+\Delta n|. \quad (15)$$

The value of $I(n,n';\Delta n)$ represents the effects of the limited overlap of the atomic probability distributions from two different initial photon numbers, as illustrated in Fig. 3b and 3c for $n = n_1$ and $n' = n_2$. This limits the coherence between photon number states $|n\rangle$ and $|n'\rangle$ to relatively close values of $n$ and $n'$, as mentioned above.

Inserting the value of $f(N_L;n)$ from Eq. (7) into Eq. (15) and converting the sum to an integral gives

$$I(n,n';\Delta n) = c_L^2 \int_{-\infty}^{\infty} dN_L e^{-(N_L-\mu(n))^2/4\sigma_L^2} e^{-(N_L-\mu(n'))^2/4\sigma_L^2} \quad (16)$$
$$\times e^{-(N_L+\Delta n-\mu(n))^2/4\sigma_L^2} e^{-(N_L+\Delta n-\mu(n'))^2/4\sigma_L^2}.$$

From Eq. (5), $\mu(n)$ is given by

$$\mu(n) = N_T P_n = N_T \varepsilon^2 n \Delta t^2 / \hbar^2. \quad (17)$$

This can be rewritten as

$$\mu(n) = \eta n, \quad (18)$$

where we have defined the constant $\eta = N_T \varepsilon^2 \Delta t^2 / \hbar^2$.

It will be assumed that the intensity in the absence of any amplification decreases as $I(z) = I_0 \exp(-\gamma z)$, where $\gamma$ is the absorption coefficient and $z$ is the distance travelled. Over a short length of fiber, the average number of absorbing atoms that make a transition to the excited state is then equal to $\bar{N}_L = n\gamma\Delta L$. With the intensity held constant instead by the distributed amplifiers over an arbitrary length $L$, we will have $\bar{N}_L = n\gamma L = \mu(n) = \eta n$ from the definition in Eq. (18). It follows that $\eta = \gamma L$ and $\eta$ is the total number of photons absorbed divided by the number of photons present initially. Equivalently, $\eta$ is the number of absorption lengths in the transmission line.

Eq. (16) can now be rewritten as

$$I(n,n';\Delta n) = c_L^2 \int_{-\infty}^{\infty} dN_L e^{-(N_L-\eta n)^2/4\sigma_L^2} e^{-(N_L-\eta n')^2/4\sigma_L^2} \quad (19)$$
$$\times e^{-(N_L+\Delta n-\eta n)^2/4\sigma_L^2} e^{-(N_L+\Delta n-\eta n')^2/4\sigma_L^2}.$$

This integral can be evaluated to give

$$I(n,n';\Delta n) = \sqrt{\pi}\sigma_L c_L^2 e^{-(n-n')^2\eta^2/4\sigma_L^2} e^{-\Delta n^2/4\sigma_L^2}. \quad (20)$$

Substituting this expression into Eq. (15) for the reduced density operator gives

$$\hat{\rho}_R = \sqrt{\pi}\sigma_L c_L^2 \sum_{\Delta n} e^{-\Delta n^2/4\sigma_L^2} \sum_{n,n'} c_n c_{n'}^* e^{-(n-n')^2\eta^2/4\sigma_L^2} \quad (21)$$
$$\times |n+\Delta n\rangle\langle n'+\Delta n|,$$

where the coefficients $c_n$ are defined in Eq. (2) as

$$c_n = \frac{1}{(2\pi n_0)^{1/4}} e^{in\phi} e^{-(n-n_0)^2/2\sigma_n^2}. \quad (22)$$

Here $n_0$ is the mean photon number in the original coherent state and $\sigma_n = \sqrt{2n_0}$.

## V. SQUEEZED COHERENT STATES

The reduced density operator of Eq. (21) could be used directly to calculate measurable parameters such as $\langle x \rangle$ or $\langle x^2 \rangle$ where $x$ is one of the field quadratures. More physical insight can be obtained, however, by introducing a state $|\tilde{\alpha}(\bar{n},\tilde{\sigma}_n)\rangle$ defined by

$$|\tilde{\alpha}(\bar{n},\tilde{\sigma}_n)\rangle = \frac{1}{\pi^{1/4}\tilde{\sigma}_n^{1/2}} \sum_{n=0}^{\infty} e^{in\phi} e^{-(n-\bar{n})^2/2\tilde{\sigma}_n^2} |n\rangle. \quad (23)$$

Eq. (23) differs from the corresponding expression for a coherent state in Eq. (2) by the replacement of $\sigma_n = \sqrt{2n_0}$ by a different standard deviation $\tilde{\sigma}_n \leq \sigma_n$. Since photon number and phase are conjugate variables, these states have a larger phase uncertainty than that of a conventional coherent state. The $|\tilde{\alpha}(\bar{n},\tilde{\sigma}_n)\rangle$ correspond to number-squeezed coherent states [21] as will be discussed in more detail in the next section.

We will now show that the reduced density operator of Eq. (21) can be rewritten in the equivalent form

$$\hat{\rho}_R = \frac{1}{2\sqrt{\pi}\tilde{\sigma}_L} \sum_{\Delta n} e^{-\Delta n^2/4\tilde{\sigma}_L^2} |\tilde{\alpha}(n_0+\Delta n,\tilde{\sigma}_n)\rangle \quad (24)$$
$$\times \langle \tilde{\alpha}(n_0+\Delta n,\tilde{\sigma}_n)|.$$

Here the constants $\tilde{\sigma}_n$ and $\tilde{\sigma}_L$ must be chosen in such a way that they satisfy the conditions

$$4\tilde{\sigma}_L^2 + \tilde{\sigma}_n^2 = 4\sigma_L^2 + \sigma_n^2 \qquad (25)$$

$$\frac{\tilde{\sigma}_L^2}{\tilde{\sigma}_n^2} = \frac{4\sigma_L^4 + 4\eta^2 \sigma_L^2 \sigma_n^2 + \eta^2 \sigma_n^4}{4\sigma_L^2 \sigma_n^2}.$$

The reduced density operator of Eq. (24) corresponds to a mixture of squeezed coherent states. Each term in the mixture has a decreased uncertainty in photon number, but the incoherent sum over $\Delta n$ increases the overall photon number uncertainty. Thus the uncertainties in both the phase and photon number are increased and the field is no longer in a minimum uncertainty state.

Eqs. (21) and (24) can be shown to be equivalent by inserting the definition of $|\tilde{\alpha}(\bar{n},\tilde{\sigma})\rangle$ into Eq. (24) to obtain

$$\hat{\rho}_R = \frac{1}{2\pi\tilde{\sigma}_L \tilde{\sigma}_n} \sum_{\Delta n} e^{-\Delta n^2 / 4\tilde{\sigma}_L^2} \sum_{n,n'}^{\infty} e^{i(n-n')\phi} e^{-(n-n_0)^2 / 2\tilde{\sigma}_n^2} e^{-(n'-n_0)^2 / 2\tilde{\sigma}_n^2}$$
$$\times |n+\Delta n\rangle\langle n'+\Delta n| \qquad (26)$$

By converting the sum over $\Delta n$ to an integral in Eqs. (21) and (26) and comparing their matrix elements in the photon-number basis, it can be shown that these two forms of $\hat{\rho}_R$ are equivalent provided that Eq. (25) is satisfied.

We will primarily be interested in the limit of large distances where $\eta \gg 1$. In that limit the solution to Eq. (25) reduces to

$$\tilde{\sigma}_n = \frac{1}{\sqrt{\eta}}\sigma'_n$$
$$\tilde{\sigma}_L = \sigma_L = \sqrt{\eta}\sigma'_n. \qquad (27)$$

Here we have defined $\sigma'_n = \sqrt{n_0}$, which is the width of the original photon number probability distribution. In the opposite limit of no loss ($\eta \to 0$), $\tilde{\sigma}_L = 0$ and $\tilde{\sigma}_n = \sigma_n$ so that $\hat{\rho}_R$ in Eq. (24) corresponds to a pure coherent state.

It can be seen from Eqs. (24) and (27) that the reduced density matrix is a mixture of squeezed states, each of which has a photon number uncertainty that is reduced by a factor of $1/\sqrt{2\eta}$ compared to that of a true coherent state. As a result, the properties of the system do not depend exponentially on the transmission length as is often the case with other kinds of systems.

Caves et al. [22] recently used a different approach to show that all phase-preserving linear amplifiers are equivalent to a parametric amplifier in which the primary mode undergoes a two-mode squeezing operation. They did not, however, write the density operator as a mixture of squeezed states as was done here in Eq. (24) and their approach was focused on characterizing the statistical distribution of the added noise rather than the effects of entanglement.

## VI. PHASE UNCERTAINTY AND THE COORDINATE REPRESENTATION

Photon number and phase are conjugate variables that satisfy the uncertainty relation $\Delta n \Delta \phi \geq \pi/2$. As a result, one would expect from Eq. (27) that the squeezed coherent states described above would have a phase uncertainty that is a factor of $\sqrt{2\eta}$ larger than that of a true coherent state in the limit of large loss. This will now be shown to be the case using the coordinate representation for a single-mode field.

A single mode of the second-quantized field is mathematically equivalent to a harmonic oscillator. As a result, we can consider a coordinate representation where the wave function $\psi(q)$ is defined [23] by

$$\psi(q) \equiv \langle q | \tilde{\alpha} \rangle. \qquad (28)$$

Here q is a generalized coordinate that is proportional to one of the quadratures of the field. For a true coherent state, the generating functions of the Hermite polynomials can be used [24,25] to show that

$$\psi_\alpha(x) = \left(\frac{1}{\pi}\right)^{1/4} \exp\left\{-\frac{x^2}{2} + \frac{2x\alpha}{\sqrt{2}} - \frac{1}{2}|\alpha|^2 - \frac{1}{2}\alpha^2\right\}, \qquad (29)$$

where the coordinate $q$ has been replaced with the dimensionless parameter $x = \sqrt{\omega/\hbar}q$.

For a true coherent state, the amplitude $\alpha$ can be written in the form

$$\alpha = i\alpha_0 e^{i\phi}, \qquad (30)$$

where $\alpha_0$ and the phase $\phi_0$ are real numbers. For the case in which $\phi \ll 1$, the argument of the exponential in Eq. (29) can be expanded to first order in $\phi$ to obtain

$$\psi(x) = \frac{1}{\pi^{1/4}} e^{i\sqrt{2}\alpha_0 x} e^{i\alpha_0^2 \phi} e^{-(x-\bar{x})^2 / 2}, \qquad (31)$$

where $\bar{x} = -\sqrt{2}\alpha_0 \phi$. A plot of the real and imaginary parts of $\psi(x)$, along with its absolute value $\psi^*(x)\psi(x)$, is shown in Fig. 4 for the case in which $\alpha_0 = 10$ and $\phi = 0.05$.

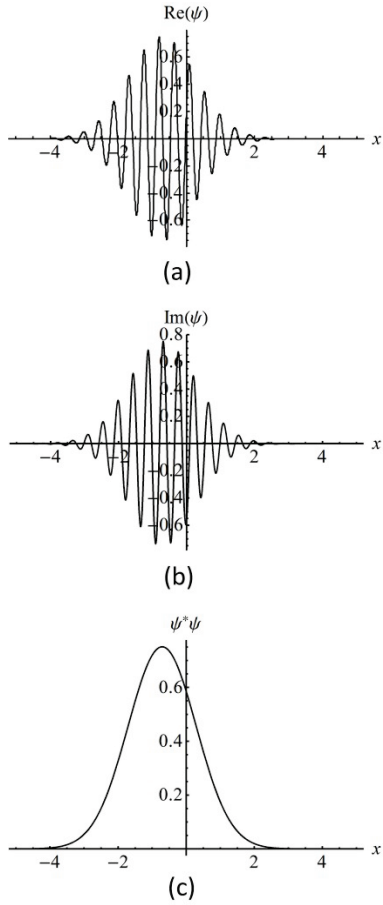

Fig. 4. Real part (a), imaginary part (b), and squared modulus (c) of the wave function $\psi(x)$ for a true coherent state in the coordinate representation. These results correspond to $\alpha_0 = 10$ and $\phi = 0.05$. (Arbitrary units.)

The fact that $|\tilde{\alpha}(\bar{n}, \tilde{\sigma})\rangle$ has the same form as a true coherent state except for the value of $\tilde{\sigma}_n < \sigma_n$ suggests that the coordinate-representation wave function $\tilde{\psi}(x)$ for a squeezed coherent state must be given approximately by

$$\tilde{\psi}(x) = \frac{1}{\pi^{1/4} \tilde{\sigma}_x^{1/2}} e^{i\sqrt{2}\alpha_0 x} e^{i\alpha_0^2 \phi} e^{-(x-\bar{x})^2 / 2 \tilde{\sigma}_x^2} \qquad (32)$$

in analogy with Eq. (31). Here $\alpha_0 = \sqrt{\bar{n}}$, $\bar{x} = -\sqrt{2}\alpha_0 \phi$, and the variance $\tilde{\sigma}_x = \sigma_n / \tilde{\sigma}_n$ reflects the increased phase uncertainty due to the reduced photon number uncertainty. Fig. 5 shows a plot of the exact value of $\tilde{\psi}(x)$ as calculated using the definition in Eq. (23) combined with the Hermite polynomial wave functions for the number states. These results correspond to a value of $\tilde{\sigma}_n = \sigma_n / 2$, with the same values of $\alpha_0$ and $\phi$ as in Fig. 4. For comparison, the corresponding values of $\tilde{\psi}(x)$ calculated using the Gaussian approximation of Eq. (32) are shown in Fig. 6. It can be seen that Eq. (32) is an excellent approximation even for relatively small values of $\alpha_0$.

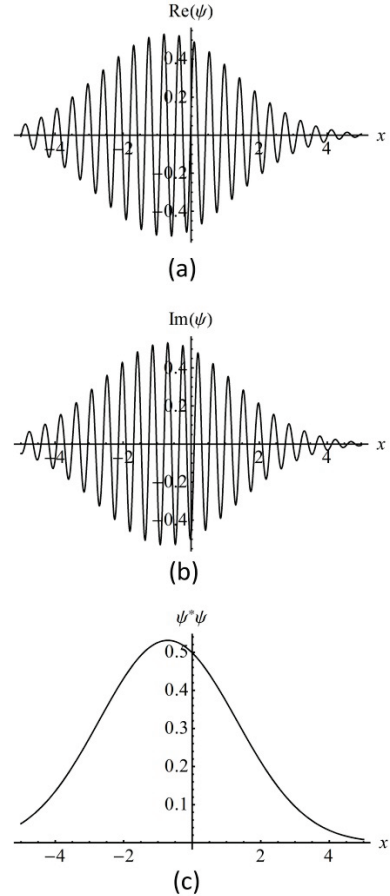

Fig. 5. Real part (a), imaginary part (b), and squared modulus (c) of the wave function $\psi(x)$ for a squeezed coherent state $|\tilde{\alpha}\rangle$ in the coordinate representation as calculated exactly using the definition in Eq. (23) and the properties of the Hermite polynomials. These results correspond to $\alpha_0 = 10$ and $\phi = 0.05$ as in Fig. 4, while $\tilde{\sigma}_n = \sigma_n / 2$. (Arbitrary units.)

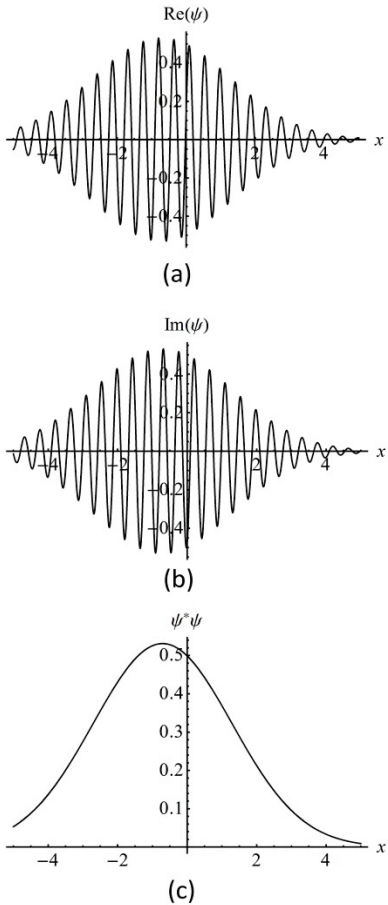

Fig. 6. Real part (a), imaginary part (b), and squared modulus (c) of the wave function $\psi(x)$ for a squeezed coherent state $|\tilde{\alpha}\rangle$ in the coordinate representation as calculated using the Gaussian approximation of Eq. (32). These results correspond to the same parameter values as in Fig. 5. (Arbitrary units.)

We will primarily be interested in homodyne phase measurements of the field quadrature $x$ under conditions in which $n_0 \gg 1$. We will assume that $\eta$ has a moderate value ($\eta \sim 10$ for example), so that there is a large amount of loss but $\tilde{\sigma}_L = \sqrt{\eta n_0} \ll n_0$. In that case $\Delta n \ll n_0$ in the density matrix of Eq. (24) and $\Delta \alpha_0 \ll \alpha_0$, where $\Delta \alpha_0$ is the typical variation in $\alpha_0 = \sqrt{n_0 + \Delta n}$ in the mixed state. We will restrict our attention to sufficiently small values of the quadrature $x$ that $\Delta \alpha_0 x \ll 1$. In that case all of the phase factors of $\exp(i\sqrt{2}\alpha_0 x)$ in Eq. (32) will be nearly independent of the value of $\Delta n$ and all of the states $|\tilde{\alpha}(n_0 + \Delta n, \tilde{\sigma}_n)\rangle$ in the mixed state will have approximately the same coordinate-representation wave function $\tilde{\psi}(x)$.

The probability amplitude to obtain a value of $x$ as a result of a homodyne measurement is just $\tilde{\psi}(x)$. Under the conditions described above, all of the $|\tilde{\alpha}(n_0 + \Delta n, \tilde{\sigma}_n)\rangle$ in the mixed state correspond to approximately the same value of $\tilde{\psi}(x)$ and will give the same results for a homodyne measurement. Thus the system can be approximately described for these purposes by a single final state $|\psi_F\rangle$ given by

$$|\psi_F\rangle = |\tilde{\alpha}(n_0, \tilde{\sigma}_n)\rangle. \tag{33}$$

The state $|\tilde{\alpha}(n_0, \tilde{\sigma}_n)\rangle$ in Eq. (33) provides an approximate description of the effects of a phase-insensitive linear distributed amplifier when $x$ is sufficiently small as described above. It is also useful in calculating the effects of decoherence on a Schrodinger cat state as described in the next section. Eq. (33) clearly neglects the effects of $\Delta n$ that would be observed if we made a measurement of photon number instead.

## VII. SCHRODINGER CAT STATES

The decoherence of Schrodinger cat states has been analyzed previously using more formal methods such as the master equation [12-17,26]. Here we will use Eq. (33) to provide an approximate description of the results of decoherence that is very simple and may provide some additional physical insight into the nature of the decoherence process.

For simplicity, consider a Schrodinger cat state that is initially given by

$$|\psi_i\rangle = \left(|\alpha_0\rangle + |e^{i\phi}\alpha_0\rangle\right), \tag{34}$$

where the constants $\phi$ and $\alpha_0$ are real and the relevant normalizing constant has been omitted. In the coordinate representation, the wave function of this superposition state can be written as

$$\psi(x) = \left(\psi_1(x) + \psi_2(x)\right). \tag{35}$$

Here $\psi_1(x)$ and $\psi_1(x)$ are given by Eq. (31) with the appropriate choice of parameters. As illustrated in Fig. 7a, the probability $P(0) = \psi^*(0)\psi(0)$ that a homodyne measurement will give the value $x = 0$ will contain an interference cross-term $T_I(x)$ given by

$$T_I(0) = 2|\psi_1(0)||\psi_2(0)|\cos(n_0\phi). \tag{36}$$

The cross-term in Eq. (36) can produce 100% visibility in the interference between the two superposed states in a Schrodinger cat in the absence of any decoherence.

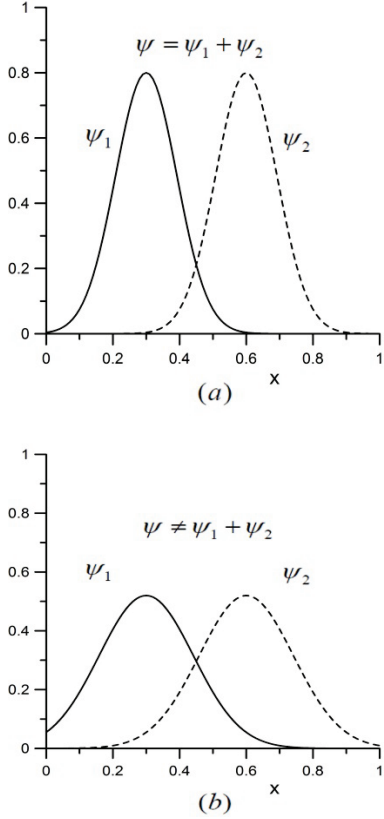

Fig. 7. Interference between the two components in a Schrodinger cat state. (a) In the absence of any loss or amplification, the initial state is a superposition of two coherent states with coordinate-representation wave functions $\psi_1(x)$ and $\psi_2(x)$, where $x$ is one of the field quadratures (proportional to the phase) . The probability amplitude for a specific value of $x$ is equal to $\psi_1(x)+\psi_2(x)$ , giving the possibility of 100% interference visibility. For simplicity, the rapidly-varying phase has not been shown here. (b) After the field has passed through a distributed amplifier, the two components in the Schrodinger cat state become entangled with different states of the environment from Eq. (39), and the probability amplitude for obtaining a specific value of $x$ is no longer simply equal to $\psi_1(x)+\psi_2(x)$ . This reduces the visibility of the interference pattern in addition to what would be expected from the phase noise. (Arbitary units.)

From Eq. (33), one might suppose that the final state $|\psi_f\rangle$ of the field after it passes through the distributed amplifier could be described approximately by

$$|\psi_f\rangle = \left(|\tilde{\alpha}_0(n_0,\tilde{\sigma}_n)\rangle + |e^{i\phi}\tilde{\alpha}_0(n_0,\tilde{\sigma}_n)\rangle\right). \quad (37)$$

If this were correct, it would produce an interference cross-term given by

$$\tilde{T}_I(0) = 2\,|\tilde{\psi}_1(0)|\,|\tilde{\psi}_2(0)|\cos(n_0\phi), \quad (38)$$

as illustrated in Fig. 7b.

But Eq. (37) neglects the fact that the two terms in the Schrodinger cat state become entangled with different states of the environment, which requires that Eq. (37) be replaced with

$$|\psi_f\rangle = \left(|\tilde{\alpha}_0(n_0,\tilde{\sigma}_n)\rangle|E_1\rangle + |e^{i\phi}\tilde{\alpha}_0(n_0,\tilde{\sigma}_n)\rangle|E_2\rangle\right). \quad (39)$$

Here $|E_1\rangle$ and $|E_2\rangle$ correspond to different states of the environment created by the passage of coherent state amplitudes differing by a phase shift $\phi$ . In that case the interference cross-term becomes

$$\tilde{T}_I(0) = 2\,|\tilde{\psi}_1(0)|\,|\tilde{\psi}_2(0)|\,|\langle E_1|E_2\rangle|\cos(n_0\phi). \quad (40)$$

It can be seen from Eq. (40) that the lack of overlap of the two environmental states will reduce the interference visibility beyond that due to the added noise.

We explicitly calculated $|E_1\rangle$ and $|E_2\rangle$ for a purely lossy channel in Ref. [25] and showed that $\langle E_1|E_2\rangle$ was in agreement with the requirements of unitarity. Here we can evaluate their overlap using the fact that the inner product of any two states must remain a constant as the system evolves. This requires that

$$\langle \alpha_0|e^{i\phi}\alpha_0\rangle = \langle \tilde{\alpha}_0(n_0,\tilde{\sigma}_n)|e^{i\phi}\tilde{\alpha}_0(n_0,\tilde{\sigma}_n)\rangle\langle E_1|E_2\rangle. \quad (41)$$

The inner products of these states can be evaluated using their wave functions in the coordinate representation and that can be used to solve for $\langle E_1|E_2\rangle$, with the result that

$$\langle E_1|E_2\rangle = \frac{\int_{-\infty}^{\infty} \psi_1^*(x)\psi_2(x)\,dx}{\int_{-\infty}^{\infty} \tilde{\psi}_1^*(x)\tilde{\psi}_2(x)\,dx}. \quad (42)$$

Eq. (42) compensates for the fact that there is an increased overlap of the wave functions in Fig. 8 as required by unitarity.

Eq. (42) can be inserted into Eq. (40) to determine the additional loss of interference visibility due to the "which path" information left in the environment by the two terms in the original superposition state. Specific examples and the application of these results to nonlocal interferometry [25,27] using macroscopic states will be discussed in a subsequent paper. It may be worth noting that some of the earliest papers on quantum noise in optical amplifiers also made use of unitarity to justify the introduction of quantum noise operators [4,6,8].

The simple form of Eq. (37) provides a straightforward way to analyze various nonlocal interference effects [25,27] when using superpositions of macroscopic coherent states.

## VIII. SUMMARY AND CONCLUSIONS

We have considered a simple model of a phase-insensitive distributed amplifier in which the electromagnetic field interacts with a series of atoms that can produce loss or gain. The state of the system including the environment (atoms) was calculated using perturbation theory. The reduced density matrix of the field was then calculated by taking the partial trace over the state of the environment. It was found that the reduced density matrix was equivalent to a mixture of number-squeezed coherent states with increased phase uncertainty. This gives a reduced density matrix with an increased uncertainty in photon number in addition to an increased phase uncertainty.

These results can be interpreted as being due to entanglement between the field and the environment as illustrated in Fig. 3. Two different photon number components $|n_1\rangle$ and $|n_2\rangle$ in the initial coherent state will be coupled to the atoms with different strengths because of the dependence of the matrix elements on the number of photons. As a result, the probability distribution $P_l(N_L;n)$ for the number $N_L$ of atoms making a transition to their excited state will be different for $n = n_1$ than it is for $n = n_2$. As the number of absorption and emission events increases, the overlap between these two probability distributions decreases and the corresponding states of the environment become nearly orthogonal for sufficiently large differences between $n_1$ and $n_2$. Although the overall uncertainty in the photon number increases due to the random-walk nature of this process, the orthogonality of the atomic states limits any coherence to relatively small differences in photon number. This results in a mixture of squeezed states $|\tilde{\alpha}(\bar{n},\tilde{\sigma}_n)\rangle$, each of which has a reduced standard deviation in photon number given by $\tilde{\sigma}_n = \sigma_n / \sqrt{2\eta}$.

The quantum noise in a phase-insensitive amplifier is often interpreted as being due to spontaneous emission noise associated with the amplification of vacuum fluctuations. It is interesting to note that vacuum fluctuations play no role in the analysis presented here. That may not be too surprising, given the fact that the probability of the field being in the vacuum state is exponentially small for $|\alpha| \gg 1$. In addition, we have used the fact that $\sqrt{n+1} \doteq \sqrt{n}$ for $|\alpha| \gg 1$, which is equivalent to neglecting spontaneous emission compared to stimulated emission. Our analysis suggests that the quantum noise and decoherence produced by a phase-insensitive distributed amplifier can be interpreted as being due to entanglement between the field and the atoms in the environment rather than vacuum fluctuations. Although $\sqrt{n+1} \doteq \sqrt{n}$ for large $n$, this is still a quantum process that would not occur for a classical field since a classical field cannot become entangled with the environment.

The entanglement with the environment and the associated decoherence can be avoided in a phase-sensitive amplifier. This has been discussed previously [2,3,5,8,12-14,16,28-31] and an analysis of phase-sensitive amplifiers is beyond the intended scope of this paper. Roughly speaking, the entanglement with the environment could be avoided in this model if the amplifying atoms were in a coherent superposition of their ground and excited states with a definite relative phase. With the correct phase of the field, the inner product between the atomic states before and after the passage of the field approaches unity and the amount of entanglement and decoherence can be minimized. Similar results can be obtained using an optical parametric amplifier.

Many previous analyses of amplifier noise were based on the introduction of a noise operator as required by unitarity or on the master equation and related techniques. For a linear amplifier, this results in a quantum noise that is added to the signal. But in addition to the added noise, decoherence can also occur as a result of entanglement between the amplifying medium and the optical field. This can be viewed as which-path information that can partially distinguish between the two components of a Schrodinger cat state, for example [25-27]. As a result, an analysis of the additive quantum noise alone is not sufficient to determine the degree of decoherence of a signal.

The approach described here includes the entanglement between the signal and the environment in a transparent way that can be readily applied to the case of Schrodinger cat states. Prior studies of decoherence of Schrodinger cat states [1-5] did not include the important case of a distributed amplifier. The results presented here can be used to analyze many systems of potential practical importance, including the effects of a distributed amplifier on long-range nonlocal interferometry using superpositions of macroscopic states [25,27].

## ACKNOWLEDGEMENTS

We would like to thank Garrett Hickman and Todd Pittman for their comments on the paper. This work was funded by grant #W31P4Q12-1-0015 from DARPA DSO.